\documentclass[conference]{IEEEtran}

\usepackage{cite}
\usepackage{graphicx}
\usepackage{amssymb, amsmath}
\usepackage{booktabs}
\usepackage{balance}
\usepackage{url}

\begin{document}

\title{Modeling Operational Fairness of Hybrid Cloud Brokerage}

\author{\IEEEauthorblockN{Sreekrishnan Venkateswaran}
\IEEEauthorblockA{IBM Corporation, India\\
Email: s\_krishna@in.ibm.com}
\and
\IEEEauthorblockN{Santonu Sarkar}
\IEEEauthorblockA{ABB Corporate Research, India\\
Email: santonus@acm.org}}

\maketitle

\begin{abstract}
Cloud service brokerage is an emerging technology that attempts to simplify the consumption and operation of hybrid clouds. Today's cloud brokers attempt to insulate consumers from the vagaries of multiple clouds. To achieve the insulation, the modern cloud broker needs to disguise itself as the end-provider to consumers by creating and operating a virtual data center construct that we call a \emph{meta-cloud}, which is assembled on top of a set of participating supplier clouds.

It is crucial for such a cloud broker to be considered a trusted partner both by cloud consumers and by the underpinning cloud suppliers. A fundamental tenet of brokerage trust is vendor neutrality. On the one hand, cloud consumers will be comfortable if a cloud broker guarantees that they will not be led through a preferred path. And on the other hand, cloud suppliers would be more interested in partnering with a cloud broker who promises a fair apportioning of client provisioning requests. Because consumer and supplier trust on a meta-cloud broker stems from the assumption of being agnostic to supplier clouds, there is a need for a test strategy that verifies the fairness of cloud brokerage.

In this paper, we propose a calculus of fairness that defines the rules to determine the operational behavior of a cloud broker. The calculus uses temporal logic to model the fact that fairness is a trait that has to be ascertained over time; it is not a characteristic that can be judged at a per-request fulfillment level. Using our temporal calculus of fairness as the basis, we propose an algorithm to determine the fairness of a broker probabilistically, based on its observed request apportioning policies. Our model for the fairness of cloud broker behavior also factors in inter-provider variables such as cost divergence and capacity variance. We empirically validate our approach by constructing a meta-cloud from AWS, Azure and IBM, in addition to leveraging a cloud simulator. Our industrial engagements with large enterprises also validate the need for such cloud brokerage with verifiable fairness.

\end{abstract}

\IEEEpeerreviewmaketitle

\section{Introduction}

Cloud service brokerage \cite{greer2013} is an emerging technology that attempts to simplify the consumption and operation of hybrid clouds. Cloud brokerage is analogous to mass-market retailers or online travel fare companies who are aggregators of services provided by end-suppliers. Cloud brokers gather multi-cloud services into a single catalog and also enable comparison across providers and a single point of purchase.

Modern cloud brokers are also seeking to shield consumers from the heterogeneity of hybrid clouds. A hybrid cloud environment provides a set of on-demand infrastructure
units with consumptive billing \cite{Assuncao2009} over which software systems can be
deployed: it can, for example, comprise of public clouds, private
clouds, as well as high-performing bare metal infrastructure. To achieve ease of hybrid consumption, the modern cloud broker can portray itself as the end-provider to consumers by composing a virtual data center abstraction that we call a \emph{meta-cloud} over participating provider clouds. Consumers will see the meta-cloud as a single cloud that offers integrated and deterministic service levels and programming interfaces, while internally leveraging the power of offerings from multiple suppliers.

It is essential for such a cloud broker to be considered a trusted partner both by cloud consumers and by underpinning cloud suppliers. While working with large global enterprise customers on designing and deploying complex brokered hybrid cloud solutions, we are increasingly discerning CIO-level apprehension and discomfort on the power and influence that a cloud broker can wield, given its vantage point on top of the entire IT (Information Technology) environment. Thus a central principle of modern brokerage is trust that stems from verifiable vendor neutrality. Because consumer and supplier trust on a meta-cloud broker spring from this assumption of vendor impartiality, there is a need for a test framework to verify the fairness of cloud brokerage, i.e., a technique to assess whether a cloud broker favors specific cloud suppliers either by design or due to inadvertent flaws. 

In this paper, we first take a high-level look at how a hybrid cloud broker can compose the meta-cloud abstraction. We then propose a system that can be used by a fairness auditor to test whether a meta-cloud broker who advertises operational fairness is indeed functioning impartially. Our work recognizes that fairness is not a characteristic that will be evident during each provisioning transaction, given that the primary priority of a broker would be to optimize the heterogeneity of a client's IT environment. Our model also realizes that the semantic of fairness is complex, given inter-cloud-provider variables such as supported infrastructure t-shirt sizes, cost, and capacity. Within these constraints, fairness needs to be a trait that will get manifest over a period across many customer interactions. We develop a calculus built on principles of temporal logic to design a framework to verify the fairness of cloud brokerage. We also back up the validity of our proposed fairness calculus with experimental results.

\subsection {Evolution of Cloud Brokerage} 
Cloud Brokerage is a technology that has recently emerged in response to expectation changes on IT consumption~\cite{marketsandmarkets}, which is seeing a fundamental shift today. Retail IT consumers and enterprise business lines now want the flexibility of choice between IT providers; they also expect simplicity and ease of usage along with the flexibility. Cloud service brokers address this by providing a uniform IT experience on top of a diverse IaaS and SaaS (Software as a Service) market place. Papers such as \cite{Iyub} have portrayed cloud brokerage as an IT supply chain construct that efficiently moves services from suppliers to consumers through the cloud network.

Cloud brokerage is an evolving technology. Cloud brokers thus far, have mainly been service aggregators; they gather multi-cloud services into a single catalog and provide a single point of purchase via a single shopping cart in a single consumption portal. The early state of cloud broker technology has thus been analogous to aggregation services offered by \emph{Expedia} \footnote{http://www.expedia.com/} in the travel industry; if an airline delivers poor service, there is no onus or liability on \emph{Expedia}. 

\begin{figure} 
\centering 
\includegraphics[scale=0.55]{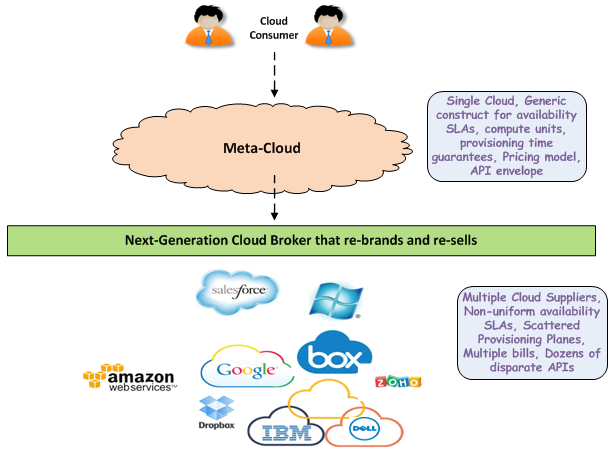}
\caption{A Brokered Meta-cloud Constructed on top of Supplier Clouds}
\label{fig:next-gen-broker}
\vspace{-10pt}
\end{figure}

However, cloud brokerage is rapidly maturing to the form that we depict in Figure~\ref{fig:next-gen-broker}, where the broker constructs and operates a virtual data center that is built on top of a heterogeneous and diverse cloud supplier marketplace \cite{greer2013}. Such a \emph{meta-cloud} offered by a cloud broker will be retail-friendly and easy to consume since it will hide complexity and deliver an integrated IT experience. A meta-cloud broker is not merely an aggregator, but a provider of cloud services over which it has control. Unlike aggregator brokers such as \emph{Expedia}, the meta-cloud broker is answerable for any violations of service levels by end-suppliers. This is analogous to the services offered by \emph{Uber} \footnote{http://www.uber.com/} in the transportation industry.

\subsection {Motivation for Verifying Fairness of Brokerage} 
In today's acquisitive market dynamics, we see an increasing number of cloud providers entering the cloud brokerage business, thereby introducing a conflict of interest. Due to this, from the perspective of commercial interest, a broker operating a meta-cloud might stand to profit by leading a customer via a preferred path. A broker could also potentially get into a secret agreement with a specific set of suppliers that give those suppliers a disproportionately high share of business as advantaged constituents of the meta-cloud. In this scenario, cloud suppliers will have a stake in ascertaining a cloud broker's trustworthiness since they would be more interested in partnering with a broker who promises a fair apportioning of client provisioning requests.

Cloud consumers will also have a stake in going with a cloud broker that guarantees that clients will not be led through a handpicked route. The brokered meta-cloud will offer value-added platform-level services in addition to being a source of infrastructure components, for example, it could provide and profit from a service that chooses a virtual machine or container from a supplier that is the best-available-fit for requested requirements. The credibility of such services will be impacted if the broker is biased to specific suppliers.

Hence, next-generation meta-cloud brokers will need to publish and promise fair behavior. It is important to note that equitable transactional behavior cannot be the unit of measurement of fairness; rather fairness is a trait expected to manifest over a period across many similar customer interactions.

This impending scenario motivated us to design and implement a fairness tester for meta-cloud brokerage.

\subsection{Related Work}
\label{sec:related-work}
To the best of our knowledge, no existing algorithms propose to verify the fairness of cloud brokers, even though there are papers such as \cite{Huang} that seek to provide a mathematical basis for general trust relationships.

\cite{Sianipar} assumes that the cloud-brokers under test are ready to co-operate with the fairness verifier in an intrusive fashion by allowing installation of auditing agents on provisioned client VMs. In many cases, this requirement will not be reasonable. Our technique does not make this simplifying assumption. The premise in \cite{Hossain} is that customers do not trust cloud providers since there could be lack of fairness in pricing and SLAs, so they want a guarantee that cloud services are refundable. The paper seeks to reduce this perceived dissatisfaction by proposing that all business procedures between the consumer and cloud providers are handled by cloud brokers; it does not deal with the question of whether the broker itself is fair.

\subsection{Organization of the Paper}
The rest of the paper is organized as follows. Section~\ref{sec:modeling-fair-brokerage} explains the semantic of what constitutes fair cloud brokerage and a model to describe a fair cloud broker. Section~\ref{sec:calculus-for-fair-brokerage} offers a logical basis for establishing fairness by proposing a calculus for fair brokerage. It starts by specifying the nature of brokered fairness using temporal logic. The inference rules that are developed in this subsection are then carried further to quantify fairness and unfairness in the form of calculus, by leveraging probability theory. Section~\ref{sec:solution-design} proposes an algorithm to determine a \emph{fairness quotient} that can be implemented by an auditor who offers fairness-testing-of-brokers-as-a-service. It also ponders on the cost complexity of implementing such a service and evaluates trade-offs between cost and accuracy. Section~\ref{sec:experimental-results} evaluates our proposed algorithm with experimental results that measure the effectiveness of our design, while Section~\ref{sec:practical-validation} describes a practical real-world validation of our work. Section~\ref{sec:validity} lists threats to validity and associated mitigation. Section\ref{sec:conclusion} highlights future research directions and concludes this paper.

\section{Modeling Fair Brokerage}
\label{sec:modeling-fair-brokerage}

\begin{figure} 
\centering 
\includegraphics[scale=0.50]{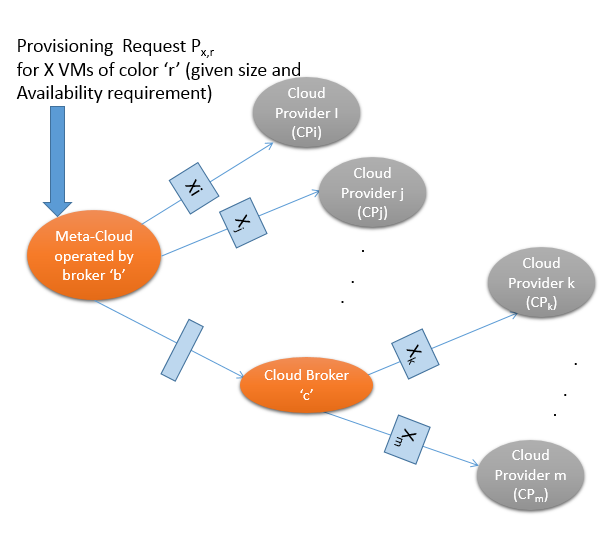}
\caption{Apportioning Provisioning Flows to Supplier Clouds}
\label{fig:apportioning-provisioning-flows}
\vspace{-10pt}
\end{figure}

In this section, we model fair composition of a brokered meta-cloud. Let $P_{x,r}$ be a provisioning request for $x$ VMs with color $r$ (where color is a combination of functional requirements such as size and non-functional requirements such as availability) received by the meta-cloud operated by cloud service broker $b$ as shown in Figure~\ref{fig:apportioning-provisioning-flows}.
Let $X_i$ be the number of VMs that the brokered meta-cloud directs to participating cloud provider $CP_i$. Then

\begin{equation}
X = \sum\limits_{i=1}^n X_i
\label{eq:pflows}
\end{equation}

where n is the number of participating cloud providers.
The problem statement that we seek to address is, \emph{for a given broker, is the apportioning of $X_i$ flows fair}?

However, fairness is not in competition with variables such as capacity and cost; rather it is complementary. To obtain such a complementary semantic of fairness, we apply the max-min fairness principle used in network traffic management\cite{Keshav} in the context of brokered provisioning. With this, we introduce \emph{efficient fairness} and interpret it as follows: Any purchase from participating cloud providers should be equitable in the long term, but consumption from a provider can be more than equitable entitlement if it does not disadvantage other providers. A cloud supplier $CP_i$ will not be disadvantaged by a distribution of $X_i$ as long as the apportionment is equitable until $CP_i$ possesses the ability to fulfill arriving provisioning requests in conformance to the required color, within cost targets. Beyond these limits of fulfillment, a provider will be deemed to have hit a flow bottleneck in the context of request $P_{x,r}$, and the concept of equitable fairness gets replaced by those mentioned above semantic of \emph{efficient fairness}.

To explain the concept of efficient fairness, let us define \emph{flow bottleneck} at an edge $i$ traversed by a provisioning flow $X_i$ as

\begin{equation}
BN(i,t,P_{x,r},b) = CT(CP_i, r, c)
\label{eq:flow-bottleneck}
\end{equation}
where\\
\begin{enumerate}
\item[$\bullet$] $BN$ is the bottleneck at edge $i$ for the provisioning flow $P_{x,r}$ that is directed to broker $b$ at time $t$,\\
\item[$\bullet$] $CT$ flags the capacity of cloud supplier $CP_i$ to provision a VM having color $r$ that does not violate the cost target $c$ (set by broker $b$). If such capacity exists, $CT$ is $1$, else it is $0$.
\end{enumerate}

If $BN(i,..)$ is 1, a bottleneck exists, so broker $b$ will not have violated \emph{efficient fairness} if provisioning flow $X_i$ no longer uses $CP_i$ as a supplier to satisfy the provisioning request being addressed. Whenever we refer to fairness relationships in subsequent sections as we develop a temporal calculus for fair brokerage, we assume this semantic of \emph{efficient fairness} expressed by a combination of Equation~\ref{eq:pflows} and Equation~\ref{eq:flow-bottleneck}.

\section{A Calculus for Fair Brokerage}
\label{sec:calculus-for-fair-brokerage}

We leverage the general concept of a calculus of trust proposed in \cite{Huang}, and develop it for our specific case of fair brokerage. We first define a fairness relationship $CurrFair (a, b, T)$, where actor $a$ evaluates $b$ and determines that it fairly performs task $T$ at the time of evaluation. We also define a complimentary relationship $CurrUnfair(a, b, T)$, where actor $a$ audits $b$ and determines that it unfairly performs task $T$ at the time of the verification. Finally, $CurrUnsure(a, b, T)$ is a relationship that depicts that, after evaluation, $a$ is unsure about the fairness or unfairness of $b$ while performing task $T$.

The outcome of each brokered provisioning request (the identity of the cloud provider to which it is directed to) resembles the roll of an unfair dice where the result of each spin can depend on the consequence of the previous one; hence the bias of the dice may change with each toss. We choose to develop a calculus based on temporal logic to model this complex scenario and predict fairness of behavior.

\subsection {Logical Basis for Broker Fairness}

Let us start by defining the terminology necessary to define a calculus for fair brokerage. Let
\begin{enumerate}
\item $f$ be the fairness tester, who is the vendor who uses our proposed algorithm to offer fairness testing of cloud brokers as a service
\item $b$ be the cloud broker whose fairness is sought to be estimated in the form of a fairness quotient
\item $c$ be the client who consumes broker-fairness-as-a-service from fairness tester $f$. This can be anyone who holds some stake in knowing whether broker $b$ is fair, for example, a prospective subscriber of meta-cloud services or a potential cloud supplier partner. 
\item $X_b$ be the task of performing brokerage with broker $b$
\item $Y_b$ be the task of performing fairness testing of cloud broker $b$ as a service
\end{enumerate}

A client $c$ can deem broker $b$ to be fair at time $t$ if
\begin{equation}
\begin{split}
CurrFair (f, b, X_b) \land CurrFair (c, f, Y_b) = True
\end{split}
\label{eq:trust-equation}
\end{equation}

This means that for a client $c$ to judge that broker $b$ operates fairly, $c$ must rate $f$ as operating fairly; and in turn, $f$ has to gauge that $b$ is performing fair brokerage. Even though fairness is a boolean decision (either "true" or "false"), the probability of a broker being fair is a number between 0 and 1. We now see how to obtain a probabilistic \emph{fairness quotient}, and how that can be converted into a logical fairness decision.

\subsection {Eventual Fairness}
\label{sec:eventual-fairness}

A broker need not apportion every request equitably since that will fragment the client deployment architecture across clouds. An efficient broker may, quite correctly, try to minimize the degree of hybrid fragmentation for customers to provide a uniform experience and to be cost sensitive. Fairness of brokerage has to thus necessarily be calculated over a period. It is possible that a broker, perceived to be unfair for a particular provisioning request or a sequence of requests, is indeed fair over a large number of independent provisioning requests. 

\begin{figure} 
\centering 
\includegraphics[scale=0.65]{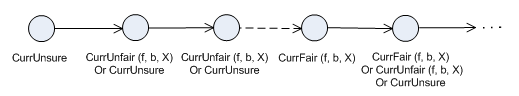}
\caption{Expected Automata of Eventual Fairness}
\label{fig:eventual-fairness}
\vspace{-10pt}
\end{figure}

This underlying nature of fairness of brokerage can be specified using Linear Temporal Logic (LTL) \cite{temporal-logic-book}, a logic that allows operational representation on a linear-time perspective. 
Let \emph{unfair} be the opposite of fair and let \emph{unsure} be the state of uncertainty, where the degree of confidence is insufficient either to determine fairness or unfairness. 
Figure~\ref{fig:eventual-fairness} depicts the intuitive meaning of fairness states. An estimate of the broker's fairness is initially in an arbitrary (or \emph{unsure}) state. As the verification algorithm starts observing the broker's operations, its state typically moves to Unfair or Unsure, given that brokered fairness is unlikely to be transactional. However, if the broker is indeed operationally fair, the verification algorithm is likely to eventually sense that fact over a period. Using temporal modalities, this property of \emph{eventual fairness} is portrayed in Figure~\ref{fig:eventual-fairness} and can be represented as:

\begin{equation}
\begin{aligned}
\lozenge CurrFair(f, b, X)
\end{aligned}
\label{eq:eventual-fairness-equation}
\end{equation}

\begin{figure} 
\centering 
\includegraphics[scale=0.65]{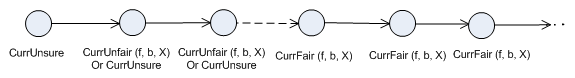}
\caption{Expected Automata of Eventual Forever Fairness}
\label{fig:eventual-forever-fairness}
\vspace{-10pt}
\end{figure}

\emph{Eventual Forever Fairness}, depicted in Figure~\ref{fig:eventual-forever-fairness}, is the scenario where the operational state remains perpetually fair, once a fair state is entered. In temporal logic terms, such overall fairness can be expressed as:

\begin{equation}
\begin{aligned}
\lozenge \square CurrFair(f, b, X)
\end{aligned}
\label{eq:eventual-forever-fairness-equation}
\end{equation}

\begin{figure} 
\centering 
\includegraphics[scale=0.65]{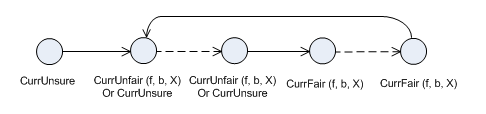}
\caption{A Real World Fair Broker in Operation ($Fair(f, b, X)$)}
\label{fig:eventual-forever-brokered-fairness}
\vspace{-10pt}
\end{figure}

A fair cloud broker in the real world, however, is unlikely to behave in this manner. A fair broker is likely to eventually reach a fair state and then move out of it. But an altruistic real world broker is likely to subsequently re-enter another fair state. Thus, each time a fair broker moves out of a fair state, there is a high chance that it eventually re-enters another fair state. We call this \emph{brokered fairness}. Figure~\ref{fig:eventual-forever-brokered-fairness} depicts this automaton in a linear-time scale while in terms of temporal logic, this property of practical brokered fairness can be represented as:
\begin{equation}
\begin{aligned}
\square \lozenge CurrFair(f, b, X)
\end{aligned}
\label{eq:eventual-forever-brokered-fairness-equation}
\end{equation}

In other words, Equation~\ref{eq:eventual-forever-brokered-fairness-equation} (and Figure~\ref{fig:eventual-forever-brokered-fairness}) represents an automata for \emph{Forever Eventual Fairness} where eventual fairness is visited infinitely often.

A broker whose operation resembles the temporal pattern expressed by Equation~\ref{eq:eventual-forever-brokered-fairness-equation} will be considered a fair broker, $Fair(f, b, X)$:
\begin{equation}
\begin{aligned}
\square \lozenge CurrFair(f, b, X) \implies Fair(f, b, X)
\end{aligned}
\label{eq:not-eventual1}
\end{equation}

When we refer to evaluated overall broker fairness in the rest of the paper, it implies $Fair(f, b, X)$ as defined in temporal terms above.

An unfair broker $Unfair(f, b, X)$, on the other hand, might accidentally encounter a fair state eventually but is unlikely to repeatedly re-enter a fair state after moving out of one. Thus, an unfair broker may sometimes encounter eventual fairness (in temporal terms, $\lnot \square \lozenge CurrFair(f, b, X)$), but not infinitely often. 

\begin{equation}
\begin{aligned}
\lnot Fair(f, b, X) \implies Unfair(f, b, X)
\end{aligned}
\label{eq:not-eventual2}
\end{equation}

In terms of the atomic predicates we defined earlier:
\begin{equation}
\begin{aligned}
\lnot \square \lozenge (\lnot(CurrUnfair(f, b, X) \vee CurrUnsure(f, b, X))) \\ 
\implies Unfair(f, b, X)
\end{aligned}
\label{eq:not-eventual3}
\end{equation}

\begin{figure} 
\centering 
\includegraphics[scale=0.58]{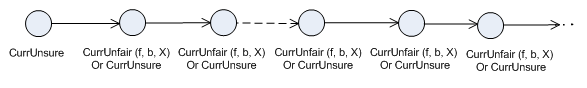}
\caption{One embodiment of an Unfair Broker in Operation}
\label{fig:not-eventual-fairness}
\vspace{-10pt}
\end{figure}

\subsection {Assessing the Broker's Fairness}
Now that we have leveraged temporal logic to develop a system of rules to represent and symbolize broker fairness, we will probabilistically quantify the above-mentioned fairness semantic in the form of calculus.

Let $\phi$, $\mu$ and $\tau$ be the probability of fair, unfair and unsure brokerage respectively. When the performer is broker $b$, we denote these symbols as $\phi_b$, $\mu_b$ and $\tau_b$. Then:
\begin{equation}
\begin{split}
\phi_b = P(Fair(f,b,X_b)=1)\\
\mu_b  = P(Fair(f,b,X_b)=0)\\
\phi_b + \mu_b + \tau_b = 1\\
\tau_b = 1 - P(Fair(f,b,X_b)=1) - P(Fair(f,b,X_b)=0)
\end{split}
\label{eq:phi-mu-tau}
\end{equation}

The odds that $f$ judges $b$ to be fair at a given point of time (one of the fair states in the automata of Figures~\ref{fig:eventual-fairness} to \ref{fig:eventual-forever-brokered-fairness}) is a probability distribution between $\phi_b$ and $\phi_b+\tau_b$ and the possibility that $f$ deems $b$ to be unfair at any point in time (one of the unfair states in the automata of Figures~\ref{fig:eventual-fairness} to \ref{fig:not-eventual-fairness}) is a distribution between $\mu_b$ and $\mu_b+\tau_b$.

Overall trust on broker fairness then translates to:
\begin{equation}
\begin{split}
Fair(f, b, X) = 1, \; if\\
                        \phi_b \geq MinFair \; \& \; \mu_b \leq MaxUnfair\\
                        or \; (\phi_b+\tau_b) \geq (1-MaxUnfair) \\ 
Fair(f, b, X) = 0    \; otherwise
\end{split}
\label{eq:more-phi-mu-tau}
\end{equation}

where $MinFair$ is the minimum acceptable probability of fair brokerage and $MaxUnfair$ is the maximum acceptable probability of unfair brokerage. 

Our algorithm sets an initial user-defined target (of $0.6$ for $MinFair$ and $0.2$ for $MaxUnfair$), but fine-tunes it over time based on an embedded feedback loop. This dynamic movement of $MinFair$ and $MaxUnfair$ is described in Section~\ref{sec:adaptive-improvements} and Section~\ref{sec:adaptive-improvements-results}. We explain how our proposed algorithm calculates $\phi_b$, $\mu_b$ and $\tau_b$ in Section~\ref{sec:probabilistic-fairness}.

\subsection {Fairness Quotient}
The \emph{fairness quotient} at time $t$ ($FQ_t$), is a metric that estimates the quantum of brokerage fairness, and as alluded to in the previous section, will be a probability distribution over $\phi_b$ to $(\phi_b + \tau_b)$. The \emph{Unfairness Quotient} at time $t$ ($UQ_t$), on the other hand, is a measure of estimated unfairness, which is a probability distribution over $\mu_b$ to $(\mu_b+\tau_b)$. For the purpose of this work, we assume that $\tau_b$ is uniformly distributed across fairness and unfairness:
\begin{equation}
\begin{aligned}
FQ_t = \phi_b + (\tau_b/2)\\
UQ_t = \mu_b + (\tau_b/2) = 1 - FQ_t\\
\end{aligned}
\label{eq:fairness-quotient-equation}
\end{equation}

\subsection {Trusting the Fairness Tester}

The client $c$ who consumes broker-fairness-as-a-service from $f$ must similarly determine the degree to which he rates $f$ as a fair service provider and the corresponding $\phi_f$, $\mu_f$ and $\tau_f$. To simplify this discussion, we assume that $c$ always considers $f$ as a fair service provider and hence:
\begin{equation}
\begin{split}
Fair(c, f, Y_b) = 1
\end{split}
\label{eq:cf-trust-equation}
\end{equation}

If we apply Equation~\ref{eq:cf-trust-equation} to Equation~\ref{eq:trust-equation}, the latter reduces to Equation~\ref{eq:more-phi-mu-tau}, which becomes the logical crux of designing a solution to estimate fairness of cloud brokerage. 

\section{Solution Design}
\label{sec:solution-design}

\subsection {Simple Fairness Test Algorithm \& Inadequacies}
\label{sec:simple-fairness}
\begin{figure} 
\centering 
\includegraphics[scale=0.30]{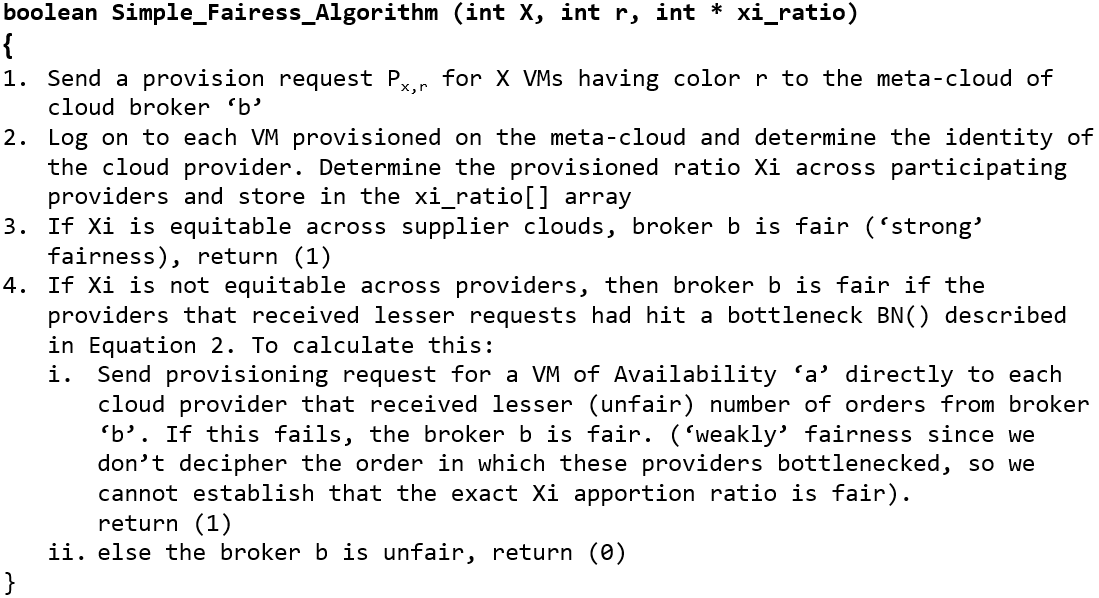}
\caption{Simple Fairness Algorithm}
\label{fig:simple-fairness-algorithm}
\vspace{-10pt}
\end{figure}

We first describe a simple algorithm to discern broker fairness. The algorithm directs provisioning requests for $X$ VMs ($X$ = $1$ to $MAX\_X$) to the brokered meta-cloud and accesses each provisioned VM to decipher the identity of the associated cloud provider. If the provisioning flow is apportioned unequally across underpinning supplier clouds, the algorithm tests if any of the providers have encountered a fulfillment bottleneck for the requested color $r$. If the apportioning is not equitable even when there are no detected bottlenecks, the broker is deemed to be unfair. The pseudo code is listed in Figure~\ref{fig:simple-fairness-algorithm}.

The following are the main inadequacies of the simple fairness test algorithm:
\begin{enumerate}
\item It discounts the possibility of eventual fairness.

\item The broker could be unfair depending on the size of the provisioning request. For example, a broker could exhibit fair behavior if the number of VMs ordered from its meta-cloud in a single provisioning request is less than $k$, and skew favorably to specific cloud providers if the provisioning request requires more than $k$ VMs. In general, a broker $b$ could be fair while fulfilling certain request sizes and unfair with others. So the simple algorithm has to be enhanced to assess fairness for each possible provisioning size. 

\item The simple algorithm is expensive (and slow) to implement and sustain since it captures the broker's behavior for all provisioning sizes as well as VM characteristics. Each provisioning request fulfilled on the brokered meta-cloud translates to cost to the provider who offers brokerage fairness verification as a service.
\end{enumerate}

This brings us to a modified and improved algorithm for verifying brokerage fairness, which we describe next.

\subsection {Modified Approach: Probabilistic Fairness Test Algorithm}
\label{sec:probabilistic-fairness}

In this section, we develop an improved algorithm that addresses the shortfalls of the simple algorithm using the temporal fairness calculus described in Section~\ref{sec:calculus-for-fair-brokerage}, and returns a predicted fairness quotient of a meta-cloud broker in addition to an opinion on whether it is operating fairly.

\begin{figure} 
\centering 
\includegraphics[scale=0.22]{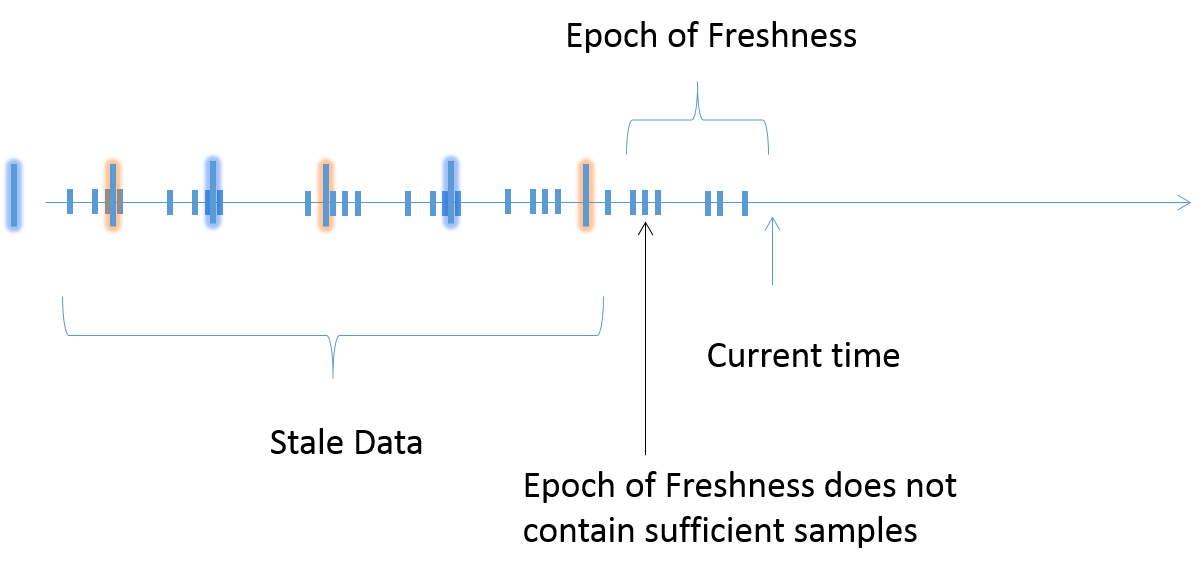}
\caption{Epoch of Freshness}
\label{fig:epoch-of-freshness}
\vspace{-10pt}
\end{figure}

\begin{figure} 
\centering 
\includegraphics[scale=0.55]{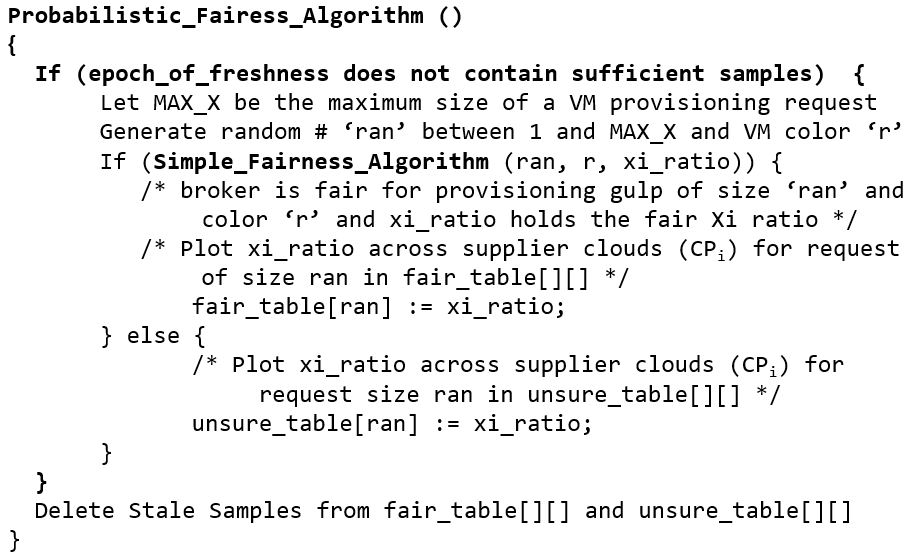}
\caption{Probabilistic Fairness Algorithm}
\label{fig:probabilistic-fairness-algorithm}
\vspace{-10pt}
\end{figure}

\begin{figure} 
\centering 
\includegraphics[scale=0.55]{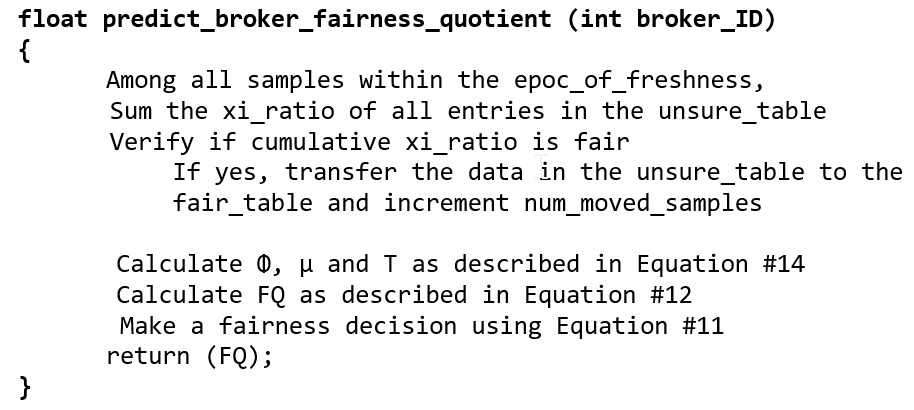}
\caption{Predicting the Fairness Quotient for a Cloud Broker}
\label{fig:predict-fairness-quotient}
\vspace{-10pt}
\end{figure}

Our modified algorithm is probabilistic and is described in Figure~\ref{fig:probabilistic-fairness-algorithm} and Figure~\ref{fig:predict-fairness-quotient}. We set an \emph{epoch of freshness}, pictorially showed in Figure~\ref{fig:epoch-of-freshness}, which is the currently running time window before which all data is considered stale. This is to recognize the fact that broker behavior, as well as supplier bottlenecks, can change over time. In each epoch, we randomly generate both the size of a provisioning request and a VM color. The size of a provisioning request falls between $1$ and $MAX\_X$, which is the maximum size of a provisioning request that the broker's meta-cloud can receive. 

\begin{figure} 
\centering 
\includegraphics[scale=0.50]{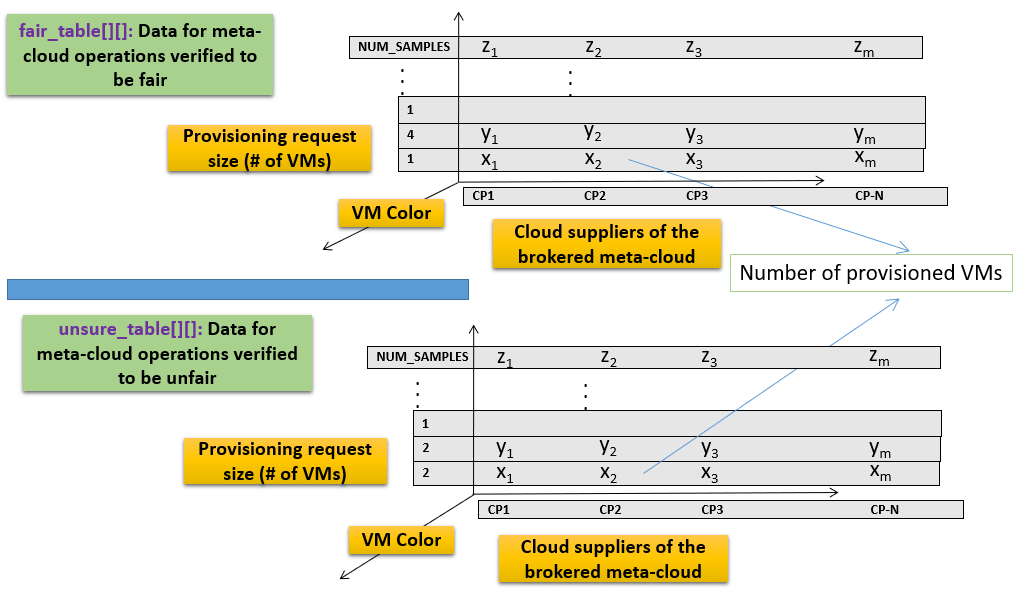}
\caption{Fairness, Unfairness and Unsure Fairness}
\label{fig:fair-unfair-table}
\vspace{-15pt}
\end{figure}

We first invoke the simple fairness algorithm to sense fairness. If deemed fair, we add the deciphered $X_i$ ratio to a table called $fair\_table$, else we add it to a table named $unsure\_table$. As shown in Figure~\ref{fig:fair-unfair-table}, these tables are three-dimensional: the $X$-axis marks the cloud providers that constitute the brokered meta-cloud, so a row in the table corresponds to an $X_i$ ratio. The $Y$-axis is the size of provisioning requests received by the meta-cloud. The Z-axis tracks the VM color.

As alluded to in Section~\ref{sec:eventual-fairness}, it must be deemed acceptable from the perspective of fairness if a broker exhibits equitable apportioning only over a period across customers. In other words, our algorithm needs the ability to detect the property of \emph{forever eventual brokered fairness}, which we depicted in Figure~\ref{fig:eventual-forever-brokered-fairness} and expressed as $Fair(f, b, X)$. To enable this, our probabilistic algorithm periodically calculates the cumulative $X_i$ ratio across $fair\_table$ and $unsure\_table$. If the accumulated $X_i$ is approximately efficiently fair, it transfers all existing entries in the $unsure\_table$ to the $fair\_table$. This is equivalent to entering an eventually brokered fair state after navigating a series of unsure states. In effect, on a linear temporal scale, the verification algorithm periodically checks whether the broker has operationally manifested fair behavior. And in course of time, it determines whether the operational automata of the broker resembles Equation~\ref{eq:eventual-forever-brokered-fairness-equation} (and Figure~\ref{fig:eventual-forever-brokered-fairness}) that echoes \emph{brokered fairness}, or whether it mirrors an automata such as the one portrayed in Equation~\ref{eq:not-eventual3} (and say, Figure~\ref{fig:not-eventual-fairness}) that translates to unfairness.

\textbf {Determining $\phi$, $\mu$ and $\tau$ to Reach Fairness Decisions:} 
The number of samples in the $fair\_table$ (called $fair\_units$) corresponds to the broker's fairness, whereas the number of samples in the $unsure\_table$ (called $unsure\_units$) corresponds to unsure fairness. As samples move from the $unsure\_table$ to the $fair\_table$, the corresponding degree of uncertainty transmutes into degree of fairness.

To extract $\tau$ from the number of samples in the $unsure\_table$, we observe the number of samples that move from the $unsure\_table$ to the $fair\_table$ in each epoch (called $moved\_units$). For a given epoch $t$, $\tau$ is calculated as the number of more samples that are expected to move during the rest of the current epoch. For this, we rely on the corresponding movement ratio across the two tables in the previous epoch, $t-1$:
\vspace{-0.03 in}
\begin{equation}
\begin{aligned}
\phi_f = fair\_units_t/(fair\_units_t + unsure\_units_t)\\
\tau_f  = 
moved\_units_{t-1} / (fair\_units_{t-1}  + unsure\_units_{t-1})\\ 
	 - moved\_units_t / (fair\_units_t + unsure\_units_t)\\
if \tau_f <0, \tau_f = 0\\
\mu_f = 1 - (\phi_f + \tau_f)\\
\end{aligned}
\label{eq:phi-mu-tau-for-fairness}
\end{equation}
A more generic way to calculate $\tau_f$ would be to trend the movement ratio across epoches with logarithmically decreasing weights for past values.

Intuitively, $\tau_f$ will be more for fair brokers than for unfair brokers.

To reach a fairness decision, we apply the results from Equation~\ref{eq:phi-mu-tau-for-fairness} to Equation~\ref{eq:more-phi-mu-tau}. If the application of Equation~\ref{eq:more-phi-mu-tau} determines the broker to be unfair, a further analysis of the $fair\_table$ and $unsure\_table$ can help determine if the unfairness is limited only to certain provisioning request sizes. After systematically eliminating the $X_i$ data corresponding to values in the $X$-axis, first in isolation and next in different combinations, if the algorithm senses fairness of operation, those provisioning request sizes that correspond to the unfair bias can be isolated.

To determine the current fairness quotient ($FQ_t$), we apply the results from Equation~\ref{eq:phi-mu-tau-for-fairness} to Equation~\ref{eq:fairness-quotient-equation}.

\subsection {Adaptive Fine-tuning of the Model's Behavior}
\label{sec:adaptive-improvements}

There are four parameters in our model that are either set by the fairness verification service provider or chosen by the service consumer:
\begin{enumerate}
\item Value for $MinFair$, the minimum acceptable probability of fair brokerage
\item Value for $MaxUnfair$, the maximum acceptable probability of unfair brokerage
\item The time length of the \emph{epoch of freshness}, the currently running time window before which all data is considered stale, and
\item The sample space size, which is the number of provisioning samples that the fairness service is allowed to gather in an epoch
\end{enumerate}

Instead of statically hard-coding these values, it is possible to dynamically re-adjust them during the operation of the fairness service. To achieve this, the fairness verification service provider will need a way to at least occasionally detect generated errors. The errors could be of two types:
\begin{enumerate}
\item \emph{False positive}, when the fairness audit marks an unfair broker as fair, and 
\item \emph{False negative}, when the audit tags a fair broker as unfair. 
\end{enumerate}

When a false positive is detected, the verification service could decrease $MaxUnfair$ in small steps, as well as gradually expand the sample space size within operational cost constraints. Conversely, when a false negative is sensed, the verification service could increase $MinFair$, as well as slightly expand the sample size. If no false positives or false negatives are encountered for a duration, the broker could dynamically increase the \emph{epoch of freshness}, as well as gradually decrease the sample size within an epoch.

\subsection {Operational Cost of the Fairness Verification Service}
\label{sec:ops-cost-versus-accuracy}

The fairness tester will incur a cost to operate. This operational cost is proportional to the number of VMs that the fairness algorithm provisions, in other words, the size of the sample space. 

\begin{figure} 
\centering 
\includegraphics[scale=0.47]{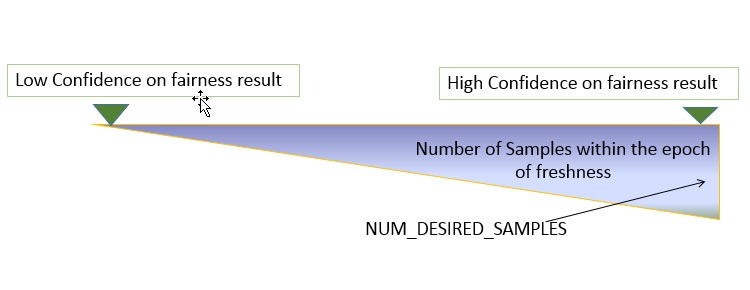}
\caption{Relation between Confidence Level and Number of Samples}
\label{fig:samples-and-confidence}
\vspace{-10pt}
\end{figure}

The accuracy of our algorithm is a function of the operational budget, i.e., the permissible cost to operate the fairness verification service. If cost is not a factor, there are no limits on the number of samples that the service can gather, so it can maintain granular data. However, in real life, target operational costs are hard constraints, so the number of permissible samples to sustain operations is limited. 
For our algorithm we set a target for the desired number of samples per epoch as shown in Figure~\ref{fig:samples-and-confidence}. The accuracy of the fairness verification service is a consequence of the targeted cost, which in turn depends on the permissible number of samples in each epoch. This aspect is reflected in the results we present in Section~\ref{sec:ops-cost-versus-accuracy-results}.

\section{Experimental Results}
\label{sec:experimental-results}

We modeled our algorithm for fairness verification by implementing 30 brokered meta-clouds, each composed of five underpinning supplier clouds - Amazon Web Services \footnote{http://aws.amazon.com/ec2/}, Microsoft Azure \footnote{http://azure.microsoft.com/}, IBM Cloud \footnote{https://www.ibm.com/cloud/} and two simulated clouds. We used the CloudSim toolkit \cite{Calheiros} to realize the two simulated clouds. Our brokers behave as follows:
\begin{enumerate}
\item For all 30 brokers, we set the total constituent capacity of all 5 supplier clouds to $1000$ VMs each, for the provisioning requests of the sole color $r$ (a combination of requirements as defined in Section~\ref{sec:modeling-fair-brokerage}) that we supported.

\item We set the duration of an epoch as 1 hour.
\item The maximum size of a provisioning request accepted by all brokered meta-clouds ($MAX\_X$) was set as $50$ VMs.
\item Half the brokers (15 out of the 30) were implemented such that they manifested fair behavior over the duration of an epoch.
\item The unfair brokers (the remaining 15) were realized such that they favored a specific supplier at a rate of between 25\% and 45\%, chosen randomly. Thus, an unfair broker that has a 25\% bias would favor a particular supplier by offering it a fourth of all received provisioning requests over an epoch. The remaining requests would be equitably distributed across all supplier clouds using the semantics of efficient fairness. This bias was distributed across sizes of provisioning requests. 

\end{enumerate}

\subsection{Predicting Broker Fairness}
\label{sec:predicting-broker-fairness-results}
We implemented the algorithm described in earlier sections to realize a broker fairness verification service. Depending on the target operational cost to implement verification, the fairness service chooses different number of samples across participating supplier clouds in each epoch. 
The outcome of our experiment is depicted in Table~\ref{tab:fair-broker-experiment-epoch-half-time}. 
The data-set shows results when the fairness service was requested half-way into the running epoch. It can be seen that while the fairness verification service indeed produced dependable results in most cases, the algorithm did not provide an accurate picture of the broker's behavior in 5 instances (Broker-5, Broker-7, Broker-10, Broker-12 and Broker-16). In general, errors can be expected when the operational cost target (and hence the sample space) is low and the unfair skew is relatively subtle.

\begin{table} [ht]
\centering
\caption{Operational Fairness Detection using our Algorithm on a Set of 30 Emulated Brokers (Service Requested Half-way into the Running Epoch)} 
\begin{tabular}{|p{0.85 in}| p{0.35 in}|p{0.53 in}|p{0.25 in}|p{0.40 in}|}
\toprule
Broker Behavior & Samples per epoch & \# of VMs Provisioned & $FQ_t$ & Fairness Decision \\
\midrule
Broker-1 (Fair) & 50 & 1246 & 0.962 & Fair \\ 
Broker-2 (Fair) & 50 & 1364 & 0.951 & Fair \\ 
Broker-3 (Fair) & 50 & 1309 & 1.0 & Fair \\ 
Broker-4 (Fair) & 50 & 1410 & 0.991 & Fair \\ 
Broker-5 (Fair) & 50 & 1334 & 0.491 & Unfair \\ 
Broker-6 (Fair) & 50 & 1201 & 0.981 & Fair \\ 
Broker-7 (Fair) & 50 & 1156 & 0.453 & Unfair \\ 
Broker-8 (Fair) & 50 & 1528 & 0.942 & Fair \\ 
Broker-9 (Fair) & 50 & 1319 & 0.938 & Fair \\ 
Broker-10 (Fair)& 50 & 1325 & 0.472 & Unfair \\ 
Broker-11 (Fair)& 50 & 1391 & 0.906 & Fair \\ 
Broker-12 (Fair)& 50 & 1350 & 0.585 & Unfair \\ 
Broker-13 (Fair)& 50 & 1050 & 0.961 & Fair \\ 
Broker-14 (Fair)& 50 & 1362 & 0.955 & Fair \\ 
Broker-15 (Fair)& 50 & 1348 & 0.962 & Fair \\ 
Broker-16 (Unfair) & 50 & 1433 & 0.992 & Fair \\ 
Broker-17 (Unfair) & 50 & 1343 & 0.226 & Unfair \\ 
Broker-18 (Unfair) & 50 & 1396 & 0.283 & Unfair \\ 
Broker-19 (Unfair) & 50 & 1246 & 0.264 & Unfair \\ 
Broker-20 (Unfair) & 50 & 1239 & 0.302 & Unfair \\ 
Broker-21 (Unfair) & 50 & 1415 & 0.321 & Unfair \\ 
Broker-22 (Unfair) & 50 & 1279 & 0.226 & Unfair \\ 
Broker-23 (Unfair) & 50 & 1176 & 0.245 & Unfair \\ 
Broker-24 (Unfair) & 50 & 1507 & 0.208 & Unfair \\ 
Broker-25 (Unfair) & 50 & 1486 & 0.302 & Unfair \\ 
Broker-26 (Unfair) & 50 & 1430 & 0.434 & Unfair \\ 
Broker-27 (Unfair) & 50 & 1525 & 0.302 & Unfair \\ 
Broker-28 (Unfair) & 50 & 1102 & 0.245 & Unfair \\ 
Broker-29 (Unfair) & 50 & 1189 & 0.245 & Unfair \\ 
Broker-30 (Unfair) & 50 & 1537 & 0.264 & Unfair \\ 
\bottomrule
\end{tabular}
\label{tab:fair-broker-experiment-epoch-half-time}
\vspace{-5pt}
\end{table}

\subsection{Adaptive Improvements}
\label{sec:adaptive-improvements-results}
In Section~\ref{sec:adaptive-improvements}, we described how a feedback loop that controls our model's key parameters can improve accuracy of the fairness analysis by reducing false positives and false negative conclusions. In this section, we back that up with results from our experiment.

For the purpose of our investigation, we chose our model's two key parameters - $MinFair$ and $MaxUnfair$ - as the variables that would be dynamically fine-tuned. We kept the the other two parameters - time length of the \emph{epoch of freshness} and the sample size within an epoch - as constants. Our adaptive improvement added the following feedback bias to the fairness evaluation algorithm when operating on the above set of 30 brokers (whose behaviors are known):
\begin{enumerate}
\item When a false positive or false negative is encountered, the algorithm turns more conservative; it narrows down the fairness window by increasing $MinFair$ and decreasing $MaxUnfair$ by one notch.
\item When a correct evaluation is reached, the algorithm turns more liberal; it broadens the fairness window by decreasing $MinFair$ and increasing $MaxUnfair$ by one notch.
\end{enumerate}

Table~\ref{tab:fair-broker-experiment-epoch-half-time-adaptive} is a rerun of Table~\ref{tab:fair-broker-experiment-epoch-half-time}, but with the above-mentioned feedback logic added to the original fairness verification algorithm. 
We now hit 3 erroneous predictions (Broker-1, Broker-4 and Broker-16) compared to 5 errors with the original algorithm realized in  Table~\ref{tab:fair-broker-experiment-epoch-half-time}. Thus, the feedback loop improved our accuracy rate; the false positive rate was cut in half.

\begin{table} [ht]
\centering
\caption{Operational Fairness Detection using our Adaptive Algorithm on a Set of 30 Emulated Brokers (Service Requested Half-way into the Running Epoch)} 
\begin{tabular}{|p{0.85 in}| p{0.35 in}|p{0.53 in}|p{0.25 in}|p{0.40 in}|}
\toprule
Broker Behavior & Samples per epoch & \# of VMs Provisioned & $FQ_t$ & Fairness Decision \\
\midrule
Broker-1 (Fair) & 50 & 1109 & 0.396 & Unfair \\ 
Broker-2 (Fair) & 50 & 1290 & 0.951 & Fair \\ 
Broker-3 (Fair) & 50 & 1323 & 0.965 & Fair \\ 
Broker-4 (Fair) & 50 & 1282 & 0.509 & Unfair \\ 
Broker-5 (Fair) & 50 & 1296 & 0.774 & Fair \\ 
Broker-6 (Fair) & 50 & 1634 & 0.981 & Fair \\ 
Broker-7 (Fair) & 50 & 1263 & 0.991 & Fair \\ 
Broker-8 (Fair) & 50 & 1317 & 0.981 & Fair \\ 
Broker-9 (Fair) & 50 & 1295 & 0.830 & Fair \\ 
Broker-10 (Fair)& 50 & 1519 & 0.981 & Fair \\ 
Broker-11 (Fair)&50 & 1241 & 0.981 & Fair \\ 
Broker-12 (Fair)& 50 & 1364 & 0.995 & Fair \\ 
Broker-13 (Fair)& 50 & 1111 & 0.990 & Fair \\ 
Broker-14 (Fair)& 50 & 1217 & 0.981 & Fair \\ 
Broker-15 (Fair)& 50 & 1260 & 0.985 & Fair \\ 
Broker-16 (Unfair) & 50 & 1450 & 0.962 & Fair \\ 
Broker-17 (Unfair) & 50 & 1285 & 0.264 & Unfair \\ 
Broker-18 (Unfair) & 50 & 1483 & 0.254 & Unfair \\ 
Broker-19 (Unfair) & 50 & 1403 & 0.169 & Unfair \\ 
Broker-20 (Unfair) & 50 & 1252 & 0.320 & Unfair \\ 
Broker-21 (Unfair) & 50 & 1395 & 0.208 & Unfair \\ 
Broker-22 (Unfair) & 50 & 1336 & 0.283 & Unfair \\ 
Broker-23 (Unfair) & 50 & 1323 & 0.132 & Unfair \\ 
Broker-24 (Unfair) & 50 & 1426 & 0.226 & Unfair \\ 
Broker-25 (Unfair) & 50 & 1328 & 0.169 & Unfair \\ 
Broker-26 (Unfair) & 50 & 1351 & 0.132 & Unfair \\ 
Broker-27 (Unfair) & 50 & 1397 & 0.339 & Unfair \\ 
Broker-28 (Unfair) & 50 & 1226 & 0.321 & Unfair \\ 
Broker-29 (Unfair) & 50 & 1162 & 0.328 & Unfair \\ 
Broker-30 (Unfair) & 50 & 1267 & 0.245 & Unfair \\ 
\bottomrule
\end{tabular}
\label{tab:fair-broker-experiment-epoch-half-time-adaptive}
\vspace{-5pt}
\end{table}

\begin{figure} 
\centering 
\includegraphics[scale=0.67]{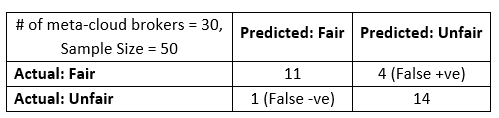}
\caption{Confusion Matrix with the Regular Algorithm}
\label{fig:confusion-non-adaptive}
\vspace{-10pt}
\end{figure}

\begin{figure} 
\centering 
\includegraphics[scale=0.67]{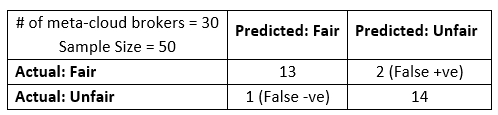}
\caption{Confusion Matrix after Introducing Adaptive Improvements}
\label{fig:confusion-adaptive}
\vspace{-10pt}
\end{figure}

We also computed a \emph{confusion matrix} with and without the adaptive improvements as shown in Figure~\ref{fig:confusion-non-adaptive} and Figure~\ref{fig:confusion-adaptive}, respectively. Figure~\ref{fig:confusion-non-adaptive} corresponds to the results contained in Table~\ref{tab:fair-broker-experiment-epoch-half-time}; Figure~\ref{fig:confusion-adaptive} corresponds to the data-set in Table~\ref{tab:fair-broker-experiment-epoch-half-time-adaptive}. The accuracy differential between the two confusion matrices establish the benefits that accrue from the refinement that we implemented.

\subsection{Operational Cost Versus Accuracy}
\label{sec:ops-cost-versus-accuracy-results}

As alluded to earlier, our fairness verification algorithm has a cost complexity that is proportional to the number of samples permitted per epoch. This, in turn, depends on the number of underpinning cloud providers who make up a broker's meta-cloud construct. 

\begin{figure} 
\centering 
\includegraphics[scale=0.75]{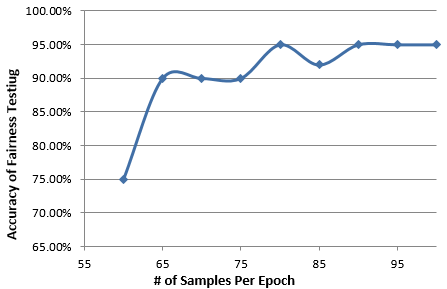}
\caption{Accuracy of Fairness Verification Versus Operational Cost}
\label{fig:cost-versus-accuracy}
\vspace{-10pt}
\end{figure}

The accuracy of the fairness assessment service should naturally be proportional to the number of permissible samples. To observe how the fairness verification algorithm's accuracy varies with operational cost, we operated our set of 30 brokers (15 fair and 15 unfair as alluded to earlier) with varying sample sizes. The results are depicted in Figure~\ref{fig:cost-versus-accuracy}. It can be observed that fairness verification accuracy plateaus at about 95\% beyond a particular sample size, but can fall to unusable levels below a certain threshold.

\section{Practical Validation}
\label{sec:practical-validation}

As part of our engagement experience in designing multi-vendor hybrid clouds for several enterprise clients, we have started encountering scenarios that highlight the importance of fairness auditing as a service. In a recent interaction with a large global bank, the CTO's team attached great importance to an overall \emph{IT Guardian} role to simplify and manage the risk of complexity inherent in their hybrid environment managed by multiple vendors.
This digitally mature customer desired to move to a model where they did not own data centers; rather they would ask the IT Guardian to host their in-house business applications with the best-fit cloud provider and offer back an on-demand consumption experience that conformed to agreed Service Level Agreements (SLAs). Generic cloud brokerage would be a key enabler in the IT Guardian operational model; see Table~\ref{tab:SIAM-provider} for the brokerage facets relevant to this role.

The bank was reluctant to consider their existing providers for the IT Guardian role due to apprehensions of potential conflict of interest. The nervousness was that IT Guardians, by virtue of the power and influence they would carry due to the advantageous position where they would reside, could have vested interest in directing more business to their own provider teams, or could lead responsibility away from their own managed service teams in the event of downtime. 

We performed the following validation of our fairness verification proposal at the IaaS pane (row \#1 of Table~\ref{tab:SIAM-provider}) with the bank's technical team: 
\begin{enumerate}
\item We verified both the relevance of the problem we are addressing in this paper and the practical acceptability of the proposed solution in the context of establishing trustworthiness of an IT Guardian.
\item We assessed and concluded that savings in total cost of ownership (TCO) will accrue if our fairness service is implemented. In the absence of a scientific and methodical fairness verification process, the selection of a guardian (that excludes all existing providers) and the learning curve for the chosen IT Guardian to be effective (given unfamiliarity of a new player with the bank's vast and complex IT environment) could consume months or even years of wasted effort.
\end{enumerate}
While we have not implemented and tested our solution in the context of this client, this could be part of future work.

\begin{table} [ht]
\centering
\caption{Trusted IT Guardian} 
\begin{tabular}{|p{1.2 in}|p{2.0 in}|}
\toprule
Facet & Relevance of Trust\\
\midrule
IaaS Broker &  Fair provisioning of infrastructure at the IaaS plane, abstracted by the meta-cloud \\ 
PaaS or SaaS Broker &  Fair placement of workloads across participating clouds \\
Service Aggregator & Fair orchestration across participating managed service providers \\
\bottomrule
\end{tabular}
\label{tab:SIAM-provider}
\vspace{-5pt}
\end{table}

\section{Threats to Validity}
\label{sec:validity}
In this section, we comment on the threats to validity while placing confidence in
our results under 3 different axes: \textit{construct validity}, \textit{internal validity}, and \textit{external validity}~\cite{Cook2002}.

Construct validity pertains to whether a metric can measure a construct adequately. Under this head, we reiterate the element of uncertain fairness that is inherent in our model and a corresponding divergence it can effect while predicting actual behavior. Given the innumerable permutations of deliberate and inadvertent unfairness with which a broker can potentially behave, and the practical limitations on the cost of testing, we cannot do away with the aspect of uncertain fairness; we can merely model and consciously handle it. Our algorithm specifically addresses this issue by treating fairness verification as a calculus, while also tracking the movement of accuracy vis-a-vis the permissible cost of ownership to operate the fairness verification service.

Internal validity highlights the impacts of the experimental process, conditions, history, or relationships among inputs on the observed outcome. While our methodology is free from overt biases, one imperfection is the challenge to choose the optimal values for $MinFair$, $MaxUndair$, the duration of the testing epoch, and the size of the sample space. To circumvent this problem, we introduced a feedback mechanism into our system that automatically and dynamically adjusts the optimal values of these variables that maximize accuracy while remaining within budgeted cost.

We have attempted to establish external validity by implementing our fairness verification algorithm, implementing a set of cloud brokers that exhibit different degrees of fairness and unfairness, and subjecting the latter to the former. We also modeled certain constraints and characteristics on top of underpinning supplier clouds. However, we have not taken into account, constructs such as dynamically priced spot instances where capacity and pricing can change even within an epoch.

\balance

\section{Conclusion and Future Work}
\label{sec:conclusion}

In this paper, we first detailed the functionality of modern cloud service brokerage and the imminent next stage of their evolution. We described how cloud brokers could monetize virtual data centers or meta-clouds constructed out of underpinning provider clouds. 

We then suggested that to engage in partnerships with cloud providers and to attract subscribers for value-added cross-cloud services on the meta-cloud, brokers will have to advertise fairness principles that guarantee impartial allocation of provisioning requests across participating provider clouds over a period. It follows that such fairness needs to be verified, so we proposed an algorithm to estimate a broker's fairness. We provided a logical basis for our method by treating fairness verification of cloud brokerage as a calculus. We showed that the nature of brokered fairness in the real world could be best represented regarding temporal logic, and hence used principles of temporal logic as the underpinning of our calculus for broker fairness. We also recognized that cost complexity of a fairness testing service must be bounded, so we designed a probabilistic algorithm that can receive operational cost as an upper bound. 

The larger topic of verifying the fairness of an IT guardian that we alluded to in Section~\ref{sec:practical-validation} will be part of our future work. With such a more generic fairness framework, any provider, even if possessing a stake in offering or managing one or more underpinning functions, can be considered to play the role of an enterprise IT guardian. More generally, the fact that an IT guardian can be verified to be fair eases the risk that enterprise IT is led through a predetermined path that directly or indirectly benefits the provider.

A logical extension of the work that we presented in this paper is to solve problems that render the meta-cloud construct rugged and commercially viable. This will include providing determinism on top of federated capacity planning and service availability. Such a verifiable deterministic meta-cloud can induce a dramatic upswing in hybrid cloud adoption due to its ease of consumption.

\section*{Acknowledgment}
We would like to thank Adarsh Srivastava, at the Birla Institute of Technology, Pilani (Goa Campus), for helping with the implementation of our fairness verification algorithm. 

\small{
\bibliographystyle{abbrv}
\bibliography{cloud.krish}
}

\end{document}